\documentstyle[aps,multicol,epsfig]{revtex}

\begin{document}

\title{Geometric features of the mixing of passive scalars at high
Schmidt numbers}
\author{J\"org Schumacher$^1$ and  Katepalli~R.~Sreenivasan$^2$}
\address{$^1$Fachbereich Physik, Philipps-Universit\"at, D-35032 Marburg,
Germany\\
$^2$International Center for Theoretical Physics, I-34014 Trieste,
Italy}
\date{\today}
\maketitle

\begin{abstract}
The mixing of passive scalars of decreasing diffusivity, advected
in each case by the same three-dimensional Navier-Stokes
turbulence, is studied. The mixing becomes more isotropic with
decreasing diffusivity. The local flow in the vicinity of steepest
negative and positive scalar gradients are in general different,
and its behavior is studied for various values of the scalar
diffusivity. Mixing approaches monofractal properties with
diminishing diffusivity. We consider these results in the context
of possible singularities of scalar dissipation.

\end{abstract}

\vspace{0.5cm}
\begin{multicols}{2}

Turbulent mixing of dyes or macromolecular substances can be
understood by analyzing how these scalar substances are advected
and diffused in a fluid medium. For small concentrations, the
admixture, or scalar, is passive, or has no dynamic feedback on
the flow---which is therefore determined independent of the
scalar. Often, the scalar diffusivity is small compared with the
viscosity of the fluid, so that their ratio, the so-called Schmidt
number, $Sc = \nu/\kappa$, is large. Our interest here lies in the
advection and diffusion of passive scalars for large values of
$Sc$. In particular, we wish to understand the recent result from
numerical simulations that turbulent mixing becomes more isotropic
when $Sc$ is increased \cite{PK02,Brethouwer}. This approach to
isotropic state is evident from the fact that deviations from it,
as quantified by odd moments of the scalar derivative along a mean
scalar gradient, $\partial_{\parallel}\theta$, decrease as $Sc$
increases from 1 to 64 for a fixed Reynolds number of the flow
\cite{PK02}.

We illustrate the issue on hand in Fig.~1, momentarily leaving
aside the details of how the data were obtained. The figure shows
the probability density function (PDF) of
$\partial_{\parallel}\theta$ for four different values of $Sc$
keeping the Reynolds number fixed. It is clear from the figure
that the distribution is quite asymmetric for $Sc = 1$ but
approaches symmetry as $Sc$ increases. This can be seen more
quantitatively in the insets. The left inset shows the quantity
$\pi(z)=[p(z)-p(-z)]/[p(z)+p(-z)]$, where $z =
\partial_{\parallel}\theta/\langle(\partial_{\parallel}\theta)^2\rangle^{1/2}$
is the normalized scalar gradient component: smaller values of this
quantity indicate smaller asymmetries. The right inset shows that
a global measure of asymmetry $A$, defined by
\begin{eqnarray}
A=\int_{-\infty}^{\infty}\,z^3 p(z) \mbox{d}z/
\int_{-\infty}^{\infty}\,|z|^3 p(z) \mbox{d}z\,,
\end{eqnarray}
also decreases with increasing $Sc$. It is easily shown
\cite{sreenivasan91} that the asymmetry in the PDF of the scalar
gradient is a measure of the anisotropy of the small-scale scalar
(or mixing), so its reduction with $Sc$ is to be regarded as a
direct indication that mixing becomes isotropic.

The reason for the asymmetry of the PDF for $Sc = O(1)$ has long
been known to be the presence of ramp-cliff structures in the
signature of the scalar field \cite{gibson} whenever there is a
mean gradient in the scalar alone or in both scalar and velocity.
This conclusion has been reinforced by recent numerical studies
\cite{Holzer,Pumir,Celani}. For the present orientation of mean
gradients, the exponential tail on the right of Fig.~1 arises
directly from cliffs in the scalar signatures, the cliffs being
related to the fronts of the scalar gradient distribution in
three-dimensional space.

Analytical approaches to understanding the features shown in
Fig.~1 are difficult. Significant progress has been made for the
so-called Kraichnan model of passive scalars \cite{Kraichnan68} in
which the advecting velocity field is assumed to be Gaussian and
to vary infinitely rapidly in time; for a review see
\cite{Gawedzki}. For two-dimensional and isotropically forced
high-$Sc$ turbulence it was possible to calculate the PDF of the
scalar dissipation rate and thus predict the scaling in the tails
of the PDF of scalar gradient fields \cite{Chertkov98}. This
Lagrangian approach was based on a sufficient separation of
pumping and diffusion time scales which is established at large
$Sc$. However, even for the synthetic Kraichnan velocity field, a
systematic study of anisotropy with $Sc$ has not been made (though
some beginnings have occurred with respect to the $Sc$-dependence
in Refs.~\cite{Balkovsky}). Another approach \cite{Schu03}, which
follows Batchelor's original model \cite{Batchelor} of quasistatic
straining motion, provides upper bounds for scalar derivative
moments as functions of $Sc$. However, the problem still contains
numerous open questions. For instance, it is unclear as to what
changes occur in and around the ramp-cliff structures as $Sc
\rightarrow \infty$. This Letter quantifies some of these changes
via numerical simulation of turbulent mixing for a range of
Schmidt numbers. We will particularly consider level sets of the
steep gradients and relate them to the local structure of the
advecting flow. The multifractal approach is known to sample
efficiently the ``singular'' structures at small scales
\cite{halsey,Sreenivasan88,Brandenburg}. Therefore, we will study
the scalar dissipation field in high-$Sc$ cases and discuss the
formation of potential singularities.

Our numerical results were obtained in a homogeneously sheared
flow \cite{Schu00} with $\langle u_x\rangle = C y$, where $C$ is a
constant, at a Taylor-microscale Reynolds number $R_{\lambda}$ of
87 in which the scalar field of constant mean gradient was allowed
to evolve according to the advection-diffusion equation. The
Schmidt number was varied from 1 to 64. The pseudospectral
simulations were done with resolutions of $512\times 257\times
512$ grid points for a box of size $2\pi:\pi:2\pi$. We ensure that
the scalar fluctuations are properly resolved by requiring that
$k_{max}\eta_B\ge 1.3$ with $k_{max}=\sqrt{2}N_{max}/3$, where
$N_{max}= 512$; the Batchelor scale $\eta_B \equiv \eta/\sqrt{Sc}$
and the Kolmogorov scale $\eta \equiv (\nu^3/\epsilon)^{1/4}$. The
scalar gradient ${\bf G}={\bf e}_y/\pi$ for all runs.

To shed some light on Fig.~1, we show in Fig.~2 three slices of
the scalar field at the same moment of evolution, each slice
corresponding to a different value of $Sc$. It is clear that the
large structure, and the front associated with it, do not change
with $Sc$ but internal striations of finer scale accompany larger
$Sc$. These striations increase the relative population of
negative gradients. This is what was observed in Fig.~1: the right
side, corresponding to positive gradients associated with fronts,
remains effectively unchanged, whereas the left side, associated
with negative gradients, shows higher probability as $Sc$
increases, thus rendering the PDFs increasingly symmetric.

The origin of the additional striations for larger $Sc$ can be
understood by writing down the evolution equation for the
derivative field, $\partial_i\theta$, where $\theta$ is the scalar
fluctuation. This equation is
\begin{eqnarray}
\left [\partial_t+{\bf u\cdot\nabla}-\kappa {\bf\nabla}^2\right]
\partial_i\theta +
\partial_i{\bf u\cdot\nabla}\theta
=-\partial_i{\bf u}{\bf\cdot G},
\end{eqnarray}
where index $i$ is kept fixed. Compared to the advection-diffusion
equation for $\theta$, the additional stretching term due to the
scalar gradient appears. This term (the last on the left hand
side) is the cause for the amplification of the scalar gradient
due to local straining motion.

The authors of Refs.~\cite{Holzer,Pumir} related
the formation of fronts to a straining motion for which the
direction of compression points into one of the scalar gradients.
Several questions arise: How robust is this result for increasing
$Sc$? What type of flow causes the steepest negative scalar
gradients, and their increase with increasing $Sc$, while the
positive gradients remain nearly unaffected? To answer these
questions, we have performed an eigenvalue analysis of the local
velocity gradients in the vicinity of the largest scalar
gradients. The starting point is the characteristic cubic
polynomial of the velocity gradient matrix
\begin{eqnarray}
\lambda^3+p \lambda+ q=0\,,
\end{eqnarray}
with
\begin{eqnarray}
p=\left[\frac{\partial(u_x,u_y)}{\partial(x,y)}+
        \frac{\partial(u_x,u_z)}{\partial(x,z)}+
        \frac{\partial(u_y,u_z)}{\partial(y,z)}\right]
\end{eqnarray}
and $q =
-\partial(u_x,u_y,u_z)/\partial(x,y,z)$.
This equation results from $\mbox{det}(\partial u_i/\partial
x_j-\lambda\delta_{ij})=0$. The solutions can be divided into
two classes. (i) For $\Delta \equiv 27q^2+4p^3 > 0$, one gets a
conjugate complex pair of eigenvalues corresponding to an inward
or outward swirl, and one real eigenvalue corresponding to the
compression or expansion in the third principal direction. (ii)
For $\Delta < 0$, all three eigenvalues are real corresponding to
pure strain. For every $Sc$, an interval in the far tails of the
scalar gradient PDF was used. Figure 3 shows, as functions of
$Sc$, the fractions of eigenvalues that were found to be purely
real or complex conjugate; panel (a) corresponds to negative tails
of the PDF and (b) to positive tails. For positive gradients, i.e.
for the cliffs, local straining becomes somewhat more dominant
with growing $Sc$, while, for the negative tails, the swirling and
straining motion contribute in almost equal parts for large $Sc$.
This is consistent with the view that the cliffs are associated
with the front stagnation point of a moving fluid parcel, where
the velocity field is predominantly straining, while the negative
slopes come from the wake region behind such parcels where the
fluid motion is both vortical and straining.

Consider now the set $F$ of all scalar concentration gradients
that exceed a certain high threshold value. From Fig.~1, we expect
the differences between the negative and positive gradients to
diminish as $Sc$ increases. The box counting dimension $D_0$ of
$F$, embedded in the three-dimensional space, is defined as the
scaling exponent of $N_{\delta}(F)=N_0 \delta^{-D_0}$ where
$N_{\delta}(F)$ is the number of cubes of size $\delta$ that are
needed to cover $F$ \cite{Falconer}. The results are shown in
Fig.~4. While, for $Sc=16$, the prefactor $N_0$ is significantly
different for the two signs of the threshold, it is about the same
for the two cases for the largest $Sc$, demonstrating that the
preference for positive gradients---corresponding to the
fronts---decreases, and that both signs of the gradient appear
with increasingly equal probability.

While keeping Sc fixed and varying the threshold we observe a
decrease of the magnitude of $D_0$ for increasing threshold (see
lower left inset of Fig.~4). The sets are thinned out and their
spatial support shrinks to lines, i.e. $D_0\simeq 1$. With
increasing threshold, the differences in $D_0$ for level sets with
opposite signs increase, which indicates that the remnants of
asymmetry for even the most symmetric case come essentially from
the far tails.

By choosing a range of level sets in the far tails of the PDF,
$|\partial_{\parallel}\theta/G|>30\pi$ for $Sc=16$, 32 and 64, we
observe that the box counting dimension decreases weakly in
magnitude for both signs of gradients (see upper right inset of
Fig.~4). Based on geometric measure theory \cite{Constantin}, or
just on the constancy of scalar flux \cite{Eyink}, it was
conjectured that the Hausdorff dimension (which is always bounded
from above by the box counting dimension) of scalar level sets for
$Sc\to\infty$ become space-filling. A gradient level set, which
stands for the change from one level of the scalar to another, can
only have spatial support between those of the two levels. If all
level sets are space-filling, the gradient level sets should show
a decreasing dimension with increasing $Sc$. This result does not
seem to apply for the present instance because we are considering
here level sets for all gradients exceeding some prescribed
(though large) threshold.

The dependence of the box-dimension on the threshold (lower left
inset of Fig.~4) is a clear evidence that the scalar derivative is
a multifractal. For operational purposes, it is more convenient to
consider the scalar dissipation field, $\epsilon_{\theta}({\bf
x},t)=\kappa \sum_{j=1}^3\,(\partial_j \theta)^2$. We define the
measure
\begin{eqnarray}
\mu^{(i)}_r=\frac{X^{(i)}_r}{X_L}= \frac{\int_{V^{(i)}_r}
\epsilon_{\theta}({\bf x}+{\bf x}_i,t) \mbox{d}^3{\bf x}}
{\int_{V} \epsilon_{\theta}({\bf x},t) \mbox{d}^3{\bf x}}
\end{eqnarray}
where $V^{(i)}_r$ is the $i$th cube with length $r$. The spectrum
of generalized dimensions \cite{Grassberger}, $D_q(q)$, obeys the
following scaling relation
\begin{eqnarray}
\sum_i \left(\mu^{(i)}_r\right)^q\sim r^{(q-1)D_q}\,.
\end{eqnarray}
In Fig.~5 the results of the multifractal analysis are shown for
$Sc=8$ and $Sc=64$. The scaling range is indicated by the dashed
lines; though it is small, it exceeds the Batchelor range and
allows the exponents to be determined with reasonable confidence.
Note that for the largest $Sc$ of 64, we have $\eta / \eta_B=8$,
i.e. less than a decade. The inset shows that the $D_q$-spectrum
gets flatter for increasing $Sc$. We have added data points from
the dye mixing experiments \cite{Sreeni90} corresponding to a
Schmidt number of about 1900.

The fact that the spectrum of generalized dimensions gets flatter
with increasing $Sc$ means that in the limit of very large $Sc$
the scalar dissipation rate becomes a monofractal, i.e. $D_q=D_0$
for all $q$. The physical picture is as follows: large
``singular'' spikes of the scalar dissipation rate fill out the
whole space more and more, so that the ``quiet'' regions
in-between become less prevalent. The degree of spatial and
temporal intermittency decreases, which is just the property to
which the spectrum is sensitive. We can quantify this result also
by fitting a bimodal multifractal cascade model to the data
\cite{Meneveau}. For $Sc=8$ we obtain $p=0.86$, for $Sc=64$ a
value of $p=0.81$, decreasing to 0.58 for the experimental data at
$Sc=1900$. The model would yield a monofractal dissipation field
for equidistributed energy flux, i.e. $p=0.5$.

We thank G.~L.~Eyink and P.K.~Yeung for useful discussions. The
computations were carried out on the IBM Blue Horizon at the San
Diego Supercomputer Center within the NPACI program of the US
National Science Foundation. KRS was supported by NSF grant
CTS-0121007.


\narrowtext

\begin{figure}
\begin{center}
\epsfig{file=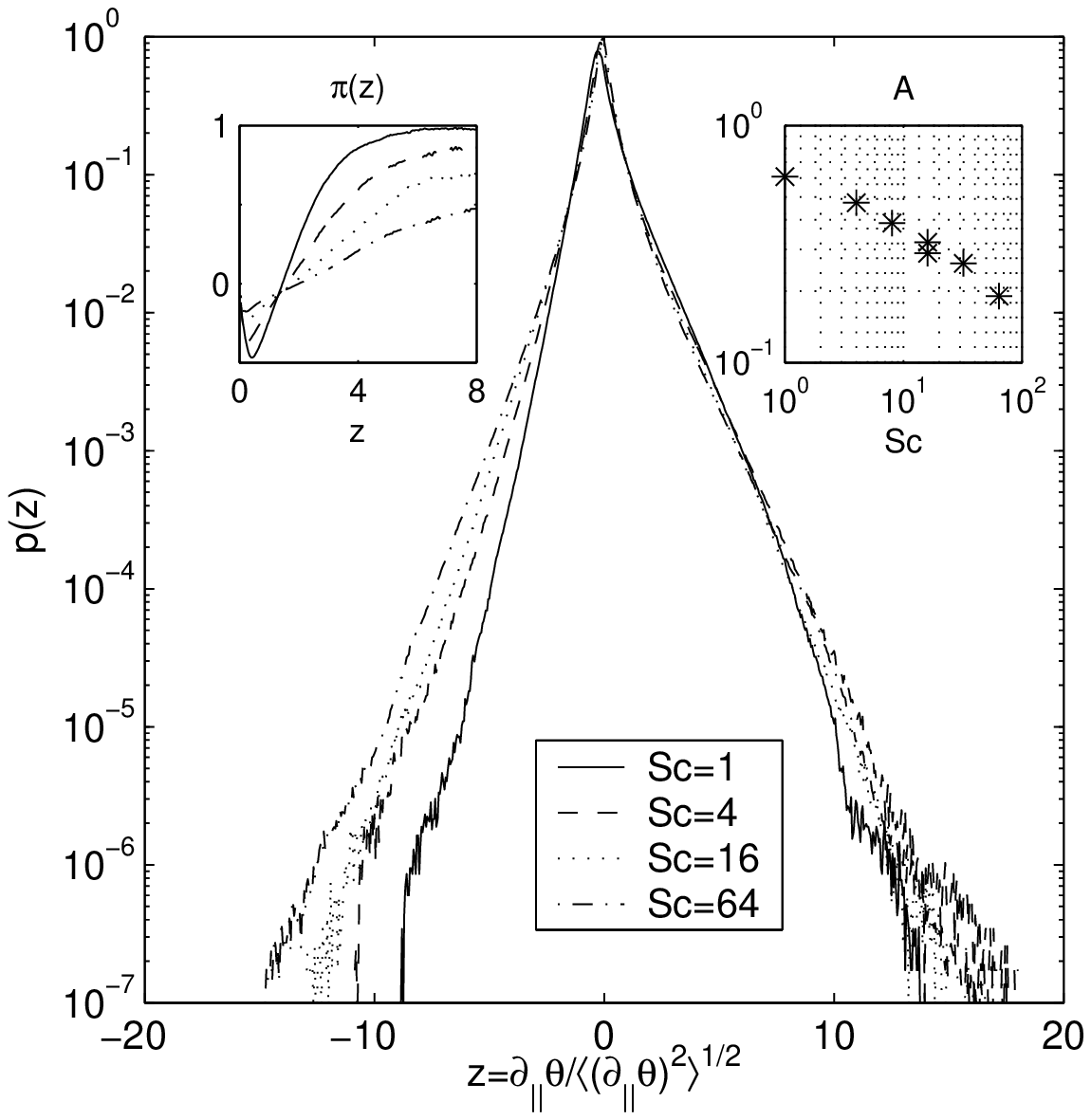,width=8cm}
\end{center}
\caption{PDFs of the normalized longitudinal scalar derivative $z$
for $Sc$ = 1, 4, 16 and 64. The scalar has a mean gradient $G$ and
is embedded in a Navier-Stokes velocity field, which is itself
sustained by a constant mean gradient, $C$. The PDFs are shown in
the main panel, and line styles in the legend. Left inset: the
ratio $\pi(z)$. Right inset: the global asymmetry factor $A$ as a
function of $Sc$. The decay of the derivative skewness, $\langle
z^3\rangle$, with increasing $Sc$ is comparable to that in [1].}
\label{fig1}
\end{figure}
\begin{figure}
\begin{center}
\epsfig{file=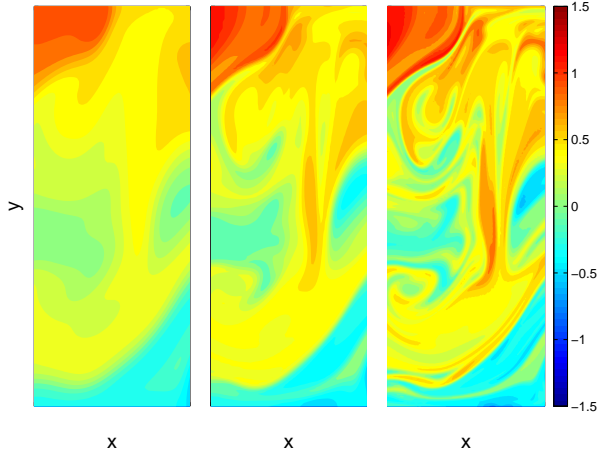,width=8cm}
\end{center}
\caption{Slices of the total scalar field (mean plus fluctuation)
for $Sc=1$ (left), 8 (middle), and 64 (right), all of which are
advected by the same flow. Only a fraction of the simulation
domain is shown.}
\label{fig2}
\end{figure}
\begin{figure}
\begin{center}
\epsfig{file=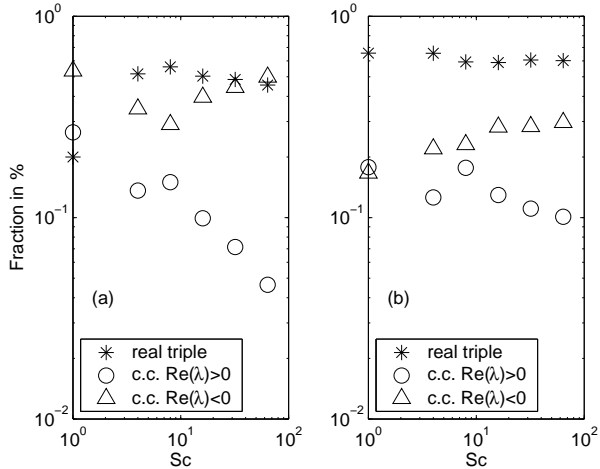,width=8cm}
\end{center}
\caption{Fractions of eigenvalue triplets, as functions of $Sc$, in
the vicinity of: (a) the steepest negative gradients, (b) steepest
positive gradients.}
\label{fig3}
\end{figure}
\begin{figure}
\begin{center}
\epsfig{file=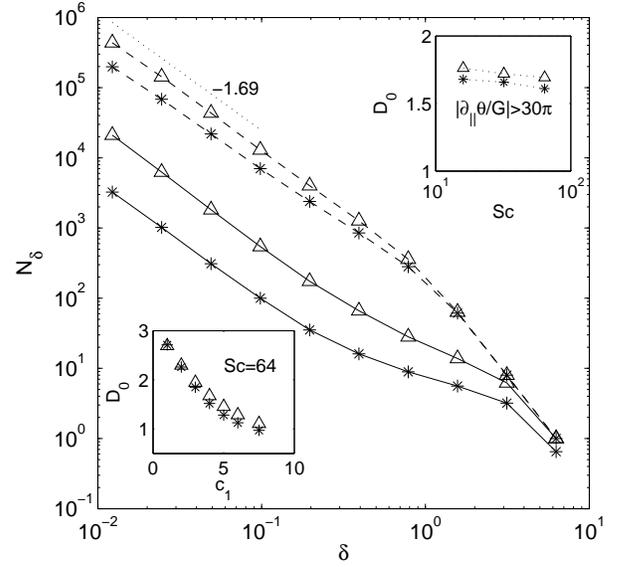,width=8cm}
\end{center}
\caption{Box counting for scalar gradient sets $F$ for $Sc$ = 16
(solid lines) and 64 (dashed lines). The sets are defined by
$|\partial_{\parallel}\theta/G|>30\pi$. Dotted lines indicate the
range of scales for the fit. Exceeding nominal expectations, the
scaling region is more extensive than the viscous-convective range
(or the Batchelor regime). Triangles: positive gradients;
asterisks: negative gradients. The differences between the
$number$ of covering boxes for positive and negative gradients are
larger for lower $Sc$ than for larger $Sc$. Upper right inset: Box
counting dimensions fitted to the data of the main figure as a
function of $Sc$. Lower left inset: Box counting dimension of a
scalar gradient level set as a function of the threshold value,
$c_1$, for $Sc=64$; here $z$ was taken within an interval $c_1\pm
0.025$. The symbols in insets are as in the main figure.}
\label{fig4}
\end{figure}
\begin{figure}
\begin{center}
\epsfig{file=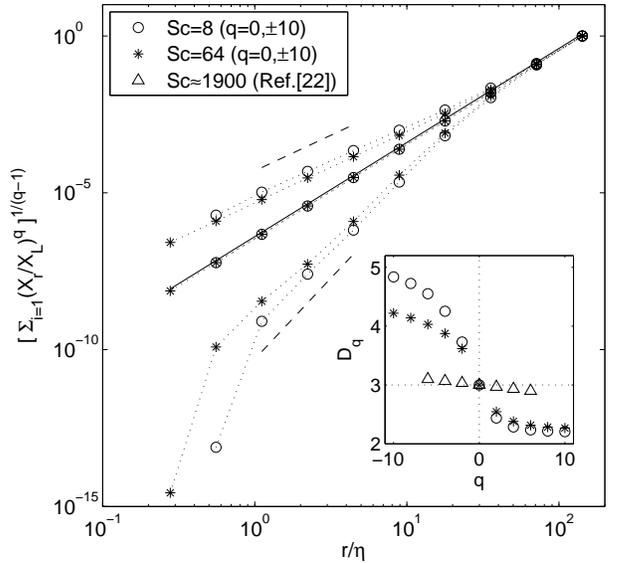,width=8cm}
\end{center}
\caption{Multifractal analysis of the scalar dissipation field for
$Sc = 8$ and 64. Because all grid points are included, we have
$D_0=3$, as indicated by the solid line. The dashed lines indicate
the scaling range used to determine the $D_q$. The inset shows the
spectrum of generalized dimensions, $D_q(q)$. Symbols for
different $Sc$ are given in the legend.}
\label{fig5}
\end{figure}
\end{multicols}

\begin{references}

\bibitem{PK02} P.~K.~Yeung, S.~Xu, and K.~R.~Sreenivasan, Phys.~Fluids {\bf 14},
               4178 (2002).

\bibitem{Brethouwer} G.~Brethouwer, J.~C.~R.~Hunt, and F.~T.~M.~Nieuwstadt,
                     J.~Fluid Mech. {\bf 474}, 193 (2003).

\bibitem{sreenivasan91} K.R.~Sreenivasan, Proc. Roy. Soc. Lond. {\bf 434}, 165
(1991).

\bibitem{gibson}
C.~H.~Gibson, C.~A.~Friehe, and S.~O.~McConnell,
              Phys.~Fluids {\bf 20}, S156 (1977);
              K.~R.~Sreenivasan, R.~A.~Antonia, and D.~Britz,
              J.~Fluid Mech. {\bf 94}, 745 (1979);
              P.~Mestayer, J.~Fluid Mech. {\bf 125}, 475 (1982);
              R.~Budwig, S.~Tavoularis, and S.~Corrsin,
              J.~Fluid Mech. {\bf 153}, 441 (1985).

\bibitem{Holzer} M.~Holzer and E.~D.~Siggia, Phys.~Fluids {\bf 6}, 1820 (1994);
                 {\it ibid} {\bf 7}, 1519 (1995).

\bibitem{Pumir} A.~Pumir, Phys.~Fluids {\bf 6}, 2118 (1994).

\bibitem{Celani} A.~Celani, A.~Lanotte, A.~Mazzino, and M.~Vergassola,
                 Phys.~Fluids {\bf 13}, 1768 (2001).

\bibitem{Kraichnan68} R.~H.~Kraichnan, Phys.~Fluids {\bf 11}, 945 (1968).

\bibitem{Gawedzki} G.~Falkovich, K.~Gaw\c{e}dzki, and M.~Vergassola,
                   Rev. Mod. Phys. {\bf 73}, 913 (2001).

\bibitem{Chertkov98} M.~Chertkov, G.~Falkovich, and I.~Kolokolov,
                     Phys. Rev. Lett. {\bf 80}, 2121 (1998).

\bibitem{Balkovsky} E.~Balkovsky and A.~Fouxon, Phys.~Rev.~E {\bf 60},
                    4164 (1999); W.~E and E.~vanden-Eijnden, Physica D
                    {\bf 152-153}, 636 (2001); M.~Chertkov and V.~Lebedev,
                    Phys.~Rev.~Lett. {\bf 90}, 034501 (2003).

\bibitem{Schu03} J.~Schumacher, K.~R.~Sreenivasan, and P.~K.~Yeung,
                 J.~Fluid~Mech. {\bf 479}, 221 (2003).

\bibitem{Batchelor} G.~K.~Batchelor, J.~Fluid~Mech. {\bf 5}, 113 (1959).

\bibitem{halsey}  T.~C.~Halsey, M.~H.~Jensen, L.~P.~Kadanoff, I.~Procaccia, and
                  B.~I.~Shraiman, Phys.~Rev. A {\bf 33}, 1141 (1986).

\bibitem{Sreenivasan88} K.~R.~Sreenivasan and C.~Meneveau, Phys.~Rev. A {\bf 38},
                        6287 (1988).

\bibitem{Brandenburg} A.~Brandenburg, I.~Procaccia, D.~Segel, and A. Vincent,
                      Phys.~Rev.~A {\bf 46}, 4819 (1992).

\bibitem{Schu00} J.~Schumacher, J. Fluid Mech. {\bf 441}, 109 (2001).

\bibitem{Falconer} K.~J.~Falconer, {\em The Geometry of Fractal Sets},
                   Cambridge University Press, Cambridge, 1986.

\bibitem{Constantin} P.~Constantin and I.~Procaccia, Nonlinearity {\bf 7},
                     1045 (1994).

\bibitem{Eyink} G.~L.~Eyink, Phys.~Rev.~E {\bf 54}, 1497 (1996).

\bibitem{Grassberger} P.~Grassberger, Phys.~Lett.~A {\bf 97}, 227 (1983);
                      H.~G.~E. Hentschel and I.~Procaccia, Physica D {\bf 8},
                      435 (1983).

\bibitem{Sreeni90} K.~R.~Sreenivasan and R.~R. Prasad, Physica D {\bf 38}, 322 (1989).

\bibitem{Meneveau} C.~Meneveau and K.~R.~Sreenivasan, Phys.~Rev.~Lett.
                   {\bf 59}, 1424 (1987).
\end{references}
\end{document}